\documentclass[reqno,11pt]{amsart}
\usepackage{fullpage,amsfonts,amssymb,amsthm,mathrsfs,graphicx,color,framed,esint}
\usepackage{cancel,bm}
\usepackage{tikz}
\usetikzlibrary{decorations.markings}
\usepackage{booktabs,longtable}

\usepackage[colorlinks=true, pdfstartview=FitV, linkcolor=blue, citecolor=blue, urlcolor=blue]{hyperref}
\definecolor{shadecolor}{rgb}{0.95, 0.95, 0.86}

\renewcommand{\d}{{\mathrm d}}

\newcommand{\im}{\mathrm{i}}
\newcommand{\e}{\mathrm{e}}

\numberwithin{equation}{section}

\newtheorem{theo}{Theorem}[section]

\begin{document}

\title[The sine process under the influence of a varying potential]{The sine process under the influence of a varying potential}

\dedicatory{To the memory of Ludwig Dmitrievich Faddeev (1934-2017)}

\author{Thomas Bothner}
\address{Department of Mathematics, University of Michigan, 2074 East Hall, 530 Church Street, Ann Arbor, MI 48109-1043, United States}
\email{bothner@umich.edu}

\author{Percy Deift}
\address{Courant Institute of Mathematical Sciences, 251 Mercer St.,New York, NY 10012, United States}
\email{deift@cims.nyu.edu}

\author{Alexander Its}
\address{Department of Mathematical Sciences, Indiana University-Purdue University Indianapolis, 402 N. Blackford St., Indianapolis, IN 46202, United States}
\email{aits@iupui.edu}

\author{Igor Krasovsky}
\address{Department of Mathematics, Imperial College, London SW7 2AZ, United Kingdom}
\email{i.krasovsky@imperial.ac.uk}



\thanks{T.B. acknowledges support of the AMS and the Simons Foundation through a travel grant. The work of P.D. is supported under NSF grant DMS-1300965 and A.I. acknowledges support of the NSF Grant DMS-1700261 and the Russian Science Foundation grant No.17-11-01126. I.K.'s work is supported by the Leverhulme Trust research project grant RPG-2018-260.}

\begin{abstract}
We review the authors' recent work \cite{BDIK1,BDIK2,BDIK3} where we obtain the uniform large $s$ asymptotics for the Fredholm determinant $D(s,\gamma):=\det(I-\gamma K_s\upharpoonright_{L^2(-1,1)})$, $0\leq\gamma\leq 1$. The operator $K_s$ acts with kernel $K_s(x,y)=\sin(s(x-y))/(\pi(x-y))$ and $D(s,\gamma)$ appears for instance in Dyson's model \cite{Dyson2} of a Coulomb log-gas with varying external potential or in the bulk scaling analysis of the thinned GUE \cite{BP}.
\end{abstract}

\date{\today}
\maketitle

\section{Introduction}
Consider the determinantal point process $\{x_j\}$ on the real line with correlation kernel
\begin{equation*}
	K(x,y):=\frac{\sin\pi(x-y)}{\pi(x-y)},\ \ \ \ x,y\in\mathbb{R}.
\end{equation*}
It is well known that this process appears in the bulk scaling limit for random matrices (in particular, in the Gaussian Unitary Ensemble GUE) where the average distance between particles $x_j$ is unity. By general theory, see \cite{S} and references therein to A. Lenard's work, for any $\phi\in L^{\infty}(\mathbb{R})$ whose support is inside a bounded measurable set $B$, 
\begin{equation}\label{e:1}
	\mathbb{E}\bigg(\prod_j\big[1-\phi(x_j)\big]\bigg)=\det(1-\phi K\upharpoonright_{L^2(B)}),
\end{equation}
where the determinant in the right-hand side of \eqref{e:1} is the Fredholm determinant of the trace-class integral operator $\phi K$ on $L^2(B)$ with kernel $\phi(x)K(x,y)$. Provided we define
\begin{equation*}
	W(x):=\begin{cases}2v,&x\in(-\frac{s}{\pi},\frac{s}{\pi})\\ 0,&\textnormal{otherwise}\end{cases},\ \ \ \ \ \textnormal{with}\ \ s,v>0,
\end{equation*}
then applying \eqref{e:1} to $\phi(x)=1-\e^{-W(x)}$ we find
\begin{equation*}
	\mathbb{E}\Big(\e^{-\sum_jW(x_j)}\Big)=\det(1-\gamma K\upharpoonright_{L^2(-\frac{s}{\pi},\frac{s}{\pi})})=\det(1-\gamma K_s\upharpoonright_{L^2(-1,1)})= D(s,\gamma),
\end{equation*}
where
\begin{equation*}
	\gamma:=1-\e^{-2v},\ \ \ \ K_s(x,y)=\frac{\sin s(x-y)}{\pi(x-y)}.
\end{equation*}
We can interpret $v>0$ above as an external potential applied to the Coulomb log-gas particle system $\{x_j\}$ whose effect it is to push particles out of the interval $(-\frac{s}{\pi},\frac{s}{\pi})$. We are interested in $D(s,\gamma)$ for large $s$ and $0\leq v\leq+\infty$. This has been a long standing problem and we shall first review relevant classical results.
\section{Asymptotic results}
\subsection{Fixed and slowly growing $v$.} If $v=+\infty$, i.e. $\gamma=1$, then $D(s,1)$ is the probability that there are no sine process particles in the interval $(-\frac{s}{\pi},\frac{s}{\pi})$. In this case,
\begin{equation}\label{e:2}
	D(s,1)=\e^{-\frac{1}{2}s^2}s^{-\frac{1}{4}}c_0\Big(1+\mathcal{O}\big(s^{-1}\big)\Big),\ \ \ \ s\rightarrow+\infty,
\end{equation}
where $\ln c_0=\frac{1}{12}\ln 2+3\zeta'(-1)$ and $\zeta'(z)$ is the derivative of Riemann's zeta function. The main term, $\e^{-\frac{1}{2}s^2}$ in \eqref{e:2} was first conjectured by Dyson, followed by a conjecture for $\e^{-\frac{1}{2}s^2}s^{-\frac{1}{4}}$ by des Cloizeaux and Mehta in \cite{dCM}. A full asymptotic expansion for $D(s,1)$, including the numerical value of $c_0$, was then identified by Dyson \cite{Dyson} in 1976 who used inverse scattering techniques for Schr\"odinger operators and an earlier work of Widom \cite{Warc} on Toeplitz determinants. Dyson's calculations were not fully rigorous and the first proof of the main term in \eqref{e:2} was given by Widom \cite{Wsine} in 1994. A proof of the full expansion (except for the value of $c_0$) was carried out in \cite{DIZ} and $c_0$, the so-called Widom-Dyson constant, finally proved in three different ways in \cite{DIKZ,Esine,Ksine}.\bigskip

If $0\leq v<s^{\frac{1}{3}}$ (which includes the case of fixed finite $v$), there exist constants $s_0,c_j>0$ such that for $s>s_0$,
\begin{equation}\label{e:3}
	D(s,\gamma)=\e^{-\frac{4v}{\pi}s}(4s)^{\frac{2v^2}{\pi^2}}G^2\Big(1+\frac{\im v}{\pi}\Big)G^2\Big(1-\frac{\im v}{\pi}\Big)\big(1+r(s,v)\big),
\end{equation}
where $G(z)$ is the Barnes G-function, 
\begin{equation*}
	G(z+1):=(2\pi)^{\frac{z}{2}}\exp\left[-\frac{z}{2}(z+1)-\frac{z^2}{2}\gamma_E\right]\prod_{k=1}^{\infty}\left(\left(1+\frac{z}{k}\right)^k\exp\left(-z+\frac{z^2}{2k}\right)\right),
\end{equation*}
with Euler's constant $\gamma_E$, and
\begin{equation*}
	\big|r(s,v)\big|<c_1\frac{v}{s}+c_2\frac{v^3}{s}.
\end{equation*}
Expansion \eqref{e:3} in the case of fixed finite $v$ as $s\rightarrow+\infty$, and with $r(s,v)$ replaced by $\mathcal{O}\big(s^{-1}\big)$ was first established by Basor and Widom in 1983 \cite{BW} and later, independently, by Budylin and Buslaev \cite{BB}. It was extended to the presented range with varying $v$ in \cite{BDIK2}. At this point it is worthwhile to contrast the Gaussian decay in the leading order of \eqref{e:2} with the leading exponential decay (for fixed $v$) in \eqref{e:3}. The underlying non-trivial transition (as worked out in the authors' works \cite{BDIK1,BDIK2,BDIK3}) is summarized in the next subsection.
\subsection{Faster growing $v$.} We shall now describe the asymptotic transition between \eqref{e:2} and \eqref{e:3} which takes place when $v$ is allowed to grow faster with $s$ than $s^{\frac{1}{3}}$. In more detail, we describe the large $s$ asymptotics of $D(s,\gamma)$ in the quarter-plane $(s,v),s,v>0$, see Figure \ref{fig1} below. Heuristically, this problem was first addressed by Dyson in \cite{Dyson2} who discovered an oscillatory behavior in the asymptotics and the appearance of Jacobi-theta functions (see \cite{BDIK1} for a discussion of Dyson's calculations). Let
\begin{equation*}
	\kappa:=\frac{v}{s}\geq 0,
\end{equation*}
so that $\gamma=1-\e^{-2v}=1-\e^{-2\kappa s}$. First, we have
\begin{theo}[\cite{BDIK1}, Corollary $1.13$ and \cite{BDIK3}]\label{res:1} As $s\rightarrow\infty$ with $\kappa>1-\frac{1}{4}\frac{\ln s}{s}$,
\begin{equation}\label{e:4}
	D(s,\gamma)=D(s,1)\big(1+o(1)\big),
\end{equation}
where the expansion for $D(s,1)$ is given in \eqref{e:2}. Moreover, when $s\rightarrow\infty$ with
\begin{equation*}
	\kappa=1-\frac{u}{2}\frac{\ln s}{s},\ \ \ \ \  q-\frac{1}{2}\leq u<q+\frac{1}{2},\ \ \ \ q=1,2,3,\ldots,\big[(\ln s)^{\frac{1}{3}}\big]+1,
\end{equation*}
where $[x]$ denotes the integer part of $x\in\mathbb{R}$,
\begin{equation}\label{e:5}
	D(s,\gamma)=D(s,1)\prod_{j=0}^{q-1}\left(1+\frac{j!}{\sqrt{\pi}}2^{-3j-2}s^{-j-\frac{1}{2}}\e^{2(s-v)}\right)\big(1+o(1)\big).
\end{equation}
\end{theo}
Observe that $\kappa=1-\frac{u}{2}\frac{\ln s}{s}$ can be written as $v=s-\frac{u}{2}\ln s$ which implies that $\e^{2(s-v)}=s^u$. Thus, if $q-\frac{1}{2}\leq u<q+\frac{1}{2}$, only the factors indexed with $j=0,1,\ldots,q-1$ contribute to the product in \eqref{e:5}, while terms with larger indices $j$ yield a decaying with $s$ contribution. In geometric terms, if one moves to the right in Figure \ref{fig1} below, each time a {\it Stokes curve} $v=s-\frac{1}{4}(2k+1)\ln s$ is crossed, a factor is added to \eqref{e:5}. One can roughly interpret this phenomenon as a particle jumping to the center of the interval $(-1,1)$, i.e. the rescaled original interval $(-\frac{s}{\pi},\frac{s}{\pi})$. Note that for $v>s-\frac{1}{4}\ln s$, estimate \eqref{e:4} shows that the leading order asymptotics of $D(s,\gamma)$ are the same as for the case $v=\infty$. The asymptotic expansions \eqref{e:4} and \eqref{e:5} in the case of fixed $q$ (rather than growing with $s$ as stated in Theorem \ref{res:1}) were proven in \cite{BDIK1} using general operator theoretical arguments and a result of Slepian \cite{Slepian} on the asymptotics of the eigenvalues of the sine-kernel integral operator. Each factor in the product \eqref{e:5} is related to an eigenvalue of the associated integral operator. In full generality, Theorem \ref{res:1} is proven in \cite{BDIK3} using Riemann-Hilbert nonlinear steepest descent techniques.\smallskip

\begin{figure}[tbh]
\resizebox{0.5\textwidth}{!}{\includegraphics{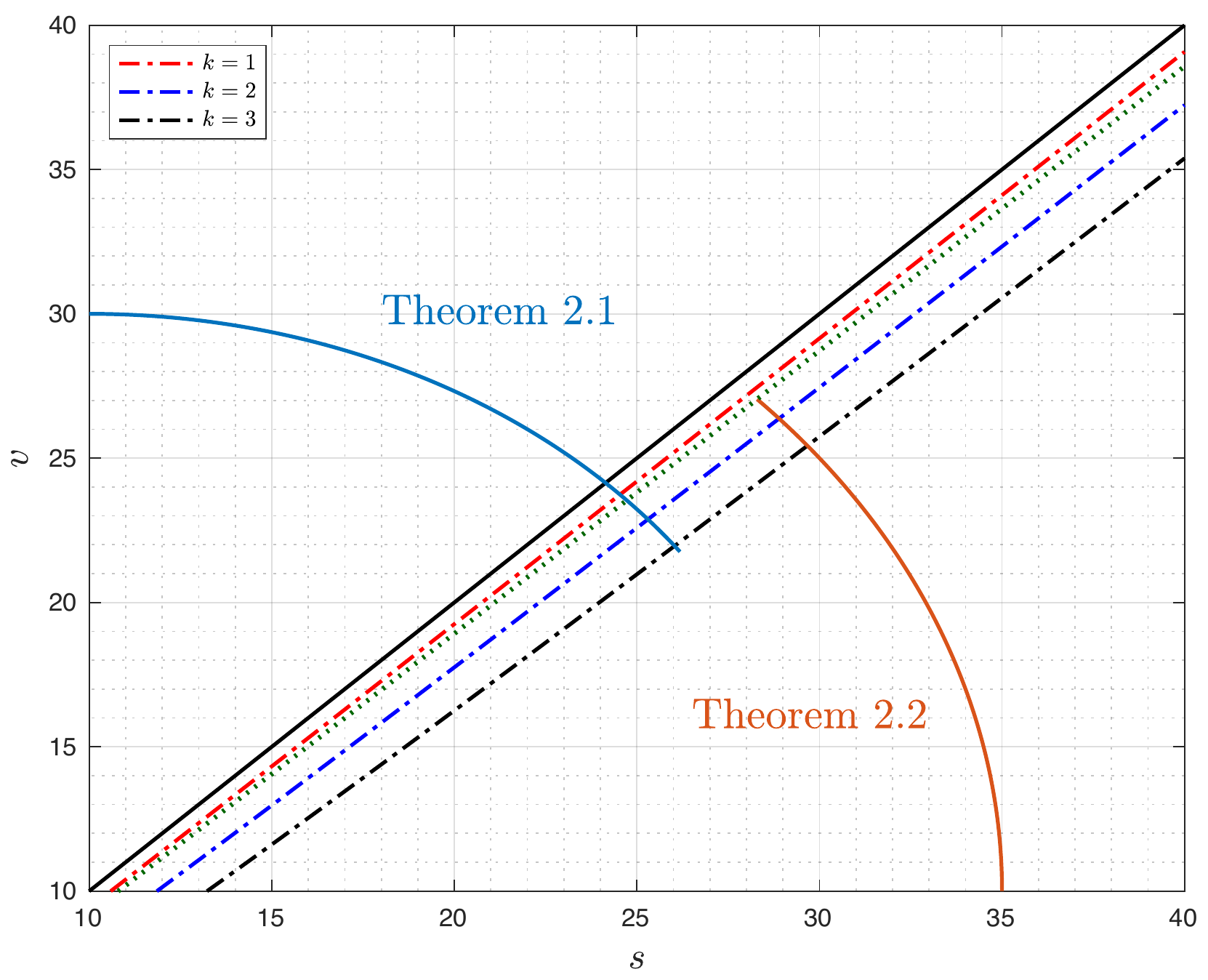}}
\caption{The quarter plane $(s,v)$ with $s,v>0$ sufficiently large. We indicate the major diagonal $v=s$ as solid black line and a few dash dotted Stokes curves $v=s-\frac{1}{4}(2k+1)\ln s$. Validity of Theorems \ref{res:1} (with $q=3$) and \ref{res:2} is indicated with two circular arcs. The dotted green curve is $v=s-\frac{1}{4}(\ln s)^{\frac{4}{3}}$.}
\label{fig1}
\end{figure}

The fact that $q$ can grow with $s$ as $[(\ln s)^{\frac{1}{3}}]$ in \eqref{e:5} (this growth estimate can be improved) allows us to connect the asymptotics of Theorem \ref{res:1} with the asymptotics in the lower region of the $(s,v)$ quarter-plane as described in the following. Let $0<a=a(\kappa)<1$ be the (unique!) solution of the equation
\begin{equation*}
	(0,1)\ni\kappa=\int_a^1\sqrt{\frac{x^2-a^2}{1-x^2}}\,\d x,
\end{equation*}
and define
\begin{equation*}
	\tau:=2\im\frac{K(a)}{K(a')},\ \ \ \ \ \ \ \ \ V:=-\frac{2}{\pi}\left[E(a)-\big(1-a^2\big)K(a)\right],
\end{equation*}
in terms of the complete elliptic integrals $K$ and $E$,
\begin{equation*}
	K(k):=\int_0^1\frac{\d t}{\sqrt{(1-t^2)(1-k^2t^2)}},\ \ \ \ \ E(k):=\int_0^1\sqrt{\frac{1-k^2t^2}{1-t^2}}\,\d t,
\end{equation*}
with modulus $a\in(0,1)$ and complementary modulus $a':=\sqrt{1-a^2}$. We also require the third Jacobi theta-function
\begin{equation*}
	\theta(z|\tau):=\sum_{k=-\infty}^{\infty}\e^{\im\pi k^2\tau+2\pi\im kz},\ \ \ \ \ -\im\tau>0,
\end{equation*}
which is periodic, $\theta(z+1|\tau)=\theta(z|\tau)$, and moreover has the property $\theta(z+\tau|\tau)=\theta(z|\tau)\e^{-\im\pi\tau-2\pi\im z}$, i.e. its second logarithmic derivative with respect to $z$ is an elliptic function with periods $1$ and $\tau$. We then have
\begin{theo}[\cite{BDIK1}, Theorem $1.4$ and \cite{BDIK3}]\label{res:2} As $s\rightarrow\infty$ with $0<\kappa<1-\frac{1}{4}\frac{(\ln s)^{\frac{4}{3}}}{s}$,
\begin{equation}\label{e:6}
	D(s,\gamma)=\e^{-\frac{1}{2}s^2(1-a^2)}\e^{vsV}\theta(sV|\tau)A(v)B(s,v)\big(1+o(1)\big),
\end{equation}
where
\begin{equation*}
	\ln A(v)=2\ln\left[G\Big(1+\frac{\im v}{\pi}\Big)G\Big(1-\frac{\im v}{\pi}\Big)\right]-\frac{v^2}{\pi^2}\Big(3-2\ln\frac{v}{\pi}\Big),
\end{equation*}
and for any $\delta>0$ there exists $C(\delta)>0$ such that $|B(s,v)|\leq C(\delta)$ for $0<\kappa<1-\delta$ as $s\rightarrow\infty$. Moreover,
\begin{equation}\label{e:7}
	\ln B(s,v)=-\frac{1}{12}\ln\big((1-\kappa)(\ln s)^5\big)+\frac{1}{6}\ln(8\pi)+o(1),\ \ \ \kappa\uparrow 1,\ s\rightarrow\infty:\ \ \ \kappa<1-\frac{1}{4}\frac{(\ln s)^{\frac{4}{3}}}{s},
\end{equation}
and
\begin{equation}\label{e:8}
	B(s,v)=1+\mathcal{O}(\kappa),\ \ \ \kappa\downarrow 0,\ s\rightarrow\infty.
\end{equation}
\end{theo}
An explicit, though somewhat cumbersome, expression for $B(s,v)$ in \eqref{e:6} in terms of Jacobi theta-functions is given in \cite{BDIK1,BDIK3}, here we are satisfied with the limiting behaviors \eqref{e:7} and \eqref{e:8}. In fact, the asymptotics of Theorem \ref{res:1} and \ref{res:2} overlap in the region
\begin{equation*}
	1-\frac{1}{2}\frac{(\ln s)^{\frac{4}{3}}}{s}<\kappa<1-\frac{1}{4}\frac{(\ln s)^{\frac{4}{3}}}{s},
\end{equation*}
where $\kappa\uparrow 1$ and thus $a\downarrow 0$. Even better, they overlap in a larger region since the estimate on $\kappa$ from above in Theorem \ref{res:2} can be somewhat increased, while the estimate on $\kappa$ from below in Theorem \ref{res:1} can be somewhat decreased. Note also that the asymptotics of Theorem \ref{res:2} are valid in the case of fixed $v$ and $s\rightarrow\infty$ when $\kappa\downarrow 0$ and $a\uparrow 1$. In this case, substituting into \eqref{e:6} the $\kappa\downarrow 0$ expansions
\begin{equation*}
	a=1-\frac{2\kappa}{\pi}-\frac{\kappa^2}{\pi^2}+\mathcal{O}\big(\kappa^3\big),\ \ \ \ \ \ V=-\frac{2}{\pi}\left(1+\frac{\kappa}{\pi}\ln\kappa-\frac{\kappa}{\pi}\big(1+\ln(4\pi)\big)+\mathcal{O}\big(\kappa^2\ln\kappa\big)\right),
\end{equation*}
and
\begin{equation*}
	\tau=-\frac{2\im}{\pi}\ln\Big(\frac{\kappa}{4\pi}\Big)+o(1),
\end{equation*}
where the last one implies that $\theta(sV|\tau)=1+o(1)$, and using \eqref{e:8}, we immediately recover \eqref{e:3}, though with a worse error estimate. Theorem \ref{res:2} above
was proven in \cite{BDIK1,BDIK2} for the narrower region $0<\kappa<1-\delta$ with $\delta$ fixed. Its proof relies on Riemann-Hilbert nonlinear steepest descent techniques which were then extended to apply to the full range of Theorem \ref{res:2} in \cite{BDIK3}.\bigskip

In summary, Theorems \ref{res:1} and \ref{res:2} provide a full analytic description of the transition between \eqref{e:2} and \eqref{e:3}. In terms of the initial log-gas particle system the following rough picture for the particles' behaviors follows from our analysis: First, for $v>s-\frac{1}{4}\ln s$, no particles are expected in the interval $(-1,1)$. As the rate of increase of $v$ with $s$ slows down, particles begin jumping one by one to the center of the interval: $q\in\mathbb{Z}_{\geq 1}$ particles at the interval center correspond to the asymptotics \eqref{e:5}, their behavior is described by classical Hermite polynomials, see \cite{BDIK3}. As $q$ increases, a transition takes place to the theta-function behavior of \eqref{e:6}, at first with $a$ close to zero. The particles then tend to fill in the centermost subinterval $(-a,a)\subset (-1,1)$ where the soft edges $\pm a$ are characterized by an approximate Airy function behavior, see \cite{BDIK1}. As the growth of $v$ with $s$ slows down further, the subinterval $(-a,a)$ increases and eventually $a$ approaches $1$. At this point a transition occurs to the limit $s\rightarrow\infty$ with $v$ fixed.
\section{Thinned sine process}
In closing, we recall the following additional interpretation of $D(s,\gamma)$ as discussed by Bohigas and Pato \cite{BP}: The Fredholm determinant $D(s,\gamma)$ equals the probability that the interval $(-\frac{s}{\pi},\frac{s}{\pi})$ contains no particles of the {\it thinned process} obtained from $\{x_j\}$ by removing each particle $x_j$ independently with likelihood $1-\gamma$. Indeed, using the well-known Fredholm determinant expression for the probability of $n$ sine-kernel process particles $x_j$ being in a given interval, we obtain via inclusion-exclusion principle
\begin{align*}
	\mathbb{P}&\big(\textnormal{no particles in}\ \Big(-\frac{s}{\pi},\frac{s}{\pi}\Big)\big)=\sum_{n=0}^{\infty}\mathbb{P}\big(n\,x_j's\ \textnormal{in}\ \Big(-\frac{s}{\pi},\frac{s}{\pi}\Big)\big)\,\mathbb{P}\big(\textnormal{these $n$ are removed}\big)\\
	=&\sum_{n=0}^{\infty}\frac{(-1)^n}{n!}\frac{\partial^n}{\partial\gamma^n}\det(1-\gamma K_s\upharpoonright_{L^2(-1,1)})\Big|_{\gamma=1}(1-\gamma)^n=\det(1-\gamma K_s\upharpoonright_{L^2(-1,1)})=D(s,\gamma).
\end{align*}
In short, $D(s,\gamma)$ is simply the gap-probability of $(-\frac{s}{\pi},\frac{s}{\pi})$ in the thinned sine-process and thus also the central building block in the evaluation of the thinned Gaudin distribtuion, i.e. the limiting distribution of gaps between consecutive thinned bulk eigenvalues. Following the arguments of Bohigas and Pato further, we contract the $s$-scale in order to keep to unity the average distance between the thinned particles, i.e. instead of $D(s,\gamma)$ we consider
\begin{equation*}
	D\Big(\frac{s}{\gamma},\gamma\Big)=\mathbb{P}\big(\textnormal{no particles in}\ \Big(-\frac{s}{\pi\gamma},\frac{s}{\pi\gamma}\Big)\big).
\end{equation*}
It follows from \eqref{e:3} that at leading order, as $\gamma=1-\e^{-2v}\downarrow 0$,
\begin{equation}\label{e:9}
	D\Big(\frac{s}{\gamma},\gamma\Big)\sim \e^{-\frac{4v}{\pi\gamma}s}\sim \e^{-\frac{2}{\pi}s},
\end{equation}
which is the gap probability of $(-\frac{s}{\pi\gamma},\frac{s}{\pi\gamma})$ in the particle system obtained from a Poisson point process by removing each particle with probability $1-\gamma$ (in order to obtain \eqref{e:9}, Bohigas and Pato used the asymptotics for fixed $v$, but a rigorous analysis requires the result for varying $\gamma$, hence varying $v$, see \cite{BDIK1}). Indeed, the gap probability in this thinned Poissonian system is exactly given by
\begin{equation*}
	\sum_{k=0}^{\infty}\frac{1}{k!}\left(\frac{2s}{\pi\gamma}\right)^k\e^{-\frac{2}{\pi\gamma}s}(1-\gamma)^k=\e^{-\frac{2}{\pi}s}.
\end{equation*}
On the other hand, for $\gamma=1$, we obtain $D(\frac{s}{\gamma},\gamma)\sim\e^{-\frac{1}{2}s^2}$, i.e. by thinning bulk eigenvalues in the GUE, we can interpolate between a Poissonian and a random matrix theory system. For other matrix ensembles, such as the Circular Unitary Ensemble CUE, corresponding gap probability phenomena have been studied in \cite{CC} and for interpolation mechanisms between random matrix theory systems and Weibull modelled particle systems we refer the interested reader to \cite{BCP,BB0,B,BIP}.

\end{document}